\documentclass{JINST}
\let\ifpdf\relax
\title{Modeling Pulse Characteristics in Xenon with NEST}
\author{J. Mock$^a$\thanks{Corresponding author}~,  N. Barry$^b$, K. Kazkaz$^c$,  D. Stolp$^a$, M. Szydagis$^a$,  M. Tripathi$^a$, S. Uvarov$^a$, M. Woods$^a$, N. Walsh$^a$\\
\llap{$^a$}University of California, Davis\\
  Department of Physics, One Shields Avenue, Davis, CA 95616, USA\\
\llap{$^b$}University of Washington\\
Department of Physics, Box 351560, Seattle, WA 98195-1560\\
\llap{$^c$}Lawrence Livermore National Laboratory\\
  7000 East Ave, Livermore, CA, 94551, USA\\
  
  E-mail: \email{jamock@ucdavis.edu}}

\abstract{A comprehensive model for describing the characteristics of pulsed signals, generated by particle interactions in xenon detectors, is presented.  An emphasis is laid on two-phase time projection chambers, but the models presented are also applicable to single phase detectors.  In order to simulate the pulse shape due to primary scintillation light,  the effects of the ratio of singlet and triplet dimer state populations, as well as their corresponding decay times, and the recombination time are incorporated into the model.  In a two phase time projection chamber, when simulating the pulse caused by  electroluminescence light, the ionization electron mean free path in gas, the drift velocity, singlet and triplet decay times, diffusion constants, and the electron trapping time, have been implemented.  This modeling has been incorporated into a complete software package, which realistically simulates the expected pulse shapes for these types of detectors.}

\keywords{Noble liquid detectors (scintillation, ionization, two-phase); Simulation methods and programs; Dark Matter detectors (WIMPs, axions, etc.); Scintillators, scintillation, and light emission processes (solid, gas, and liquid scintillators)}

\begin{document}

\section{Introduction}
Detectors employing liquid noble elements as target materials are actively involved in searching for dark matter and neutrino-less double beta decay.  There are a number of experiments underway, or in the planning stages, that use liquid xenon, neon, or argon as a detection medium \cite{Schnee2011,AprileDoke} .  Because of the non-linearity of the energy dependence of the scintillation yield, these detectors present challenges for accurate modeling and simulation.   The Noble Element Simulation Technique (NEST) was developed as an extension to Geant4 \cite{Agostinelli2003} to rectify this problem.  NEST provides a comprehensive and accurate simulation of the excitation, ionization, and recombination processes, which produce the primary scintillation and secondary electroluminescence light in liquid noble elements.  The central aspect is the modeling of recombination probability of ionization electrons, thereby correctly simulating the scintillation and charge yields.  This has been demonstrated for  both electron recoils (ER) and nuclear recoils (NR)  in liquid or gas \cite{Szydagis:2011tk}.  Here we present a collection of new models  that  simulate pulse characteristics of the primary scintillation light (S1) and the electroluminescence signal (S2)  in a dual phase xenon detector.  These models have been incorporated within the NEST simulation package\footnote{To download the package visit http://nest.physics.ucdavis.edu}.

\section{S1 Pulse Shape}
The primary scintillation signal in xenon is generated via two processes.  When an incoming particle enters the target medium, it deposits energy  through either direct atomic excitation, leading to the creation of a dimer in the excited state, or ionization, which produces free electrons.  The free electrons can either escape the interaction site or recombine within a characteristic timescale producing excited dimers, similar to those created by direct excitation. The de-excitation of all dimers produces the total S1 \cite{Kubota1979}.

The time structure of the S1 pulse shape can be characterized by three fundamental time constants: the singlet lifetime, the triplet lifetime, and the recombination time, as well as the ratio of singlet states to triplet states.  In addition, effects due to detector geometry and signal processing add further modifications to the pulse shape.  These are not the subject of NEST models and should be incorporated into user-specific Geant4 simulations. Here we focus on the physical time constants included in S1 pulse shape models for xenon, as summarized  in Table \ref{table:table1}.

\begin{table}
\caption{Parameters that affect the shape of the S1 pulse in liquid xenon detectors.  The lifetimes for the singlet and triplet dimer states are listed.
The ratios of the number of excitations in the singlet state to those in the triplet state are also listed for various cases. ER events corresponding to direct excitation or recombination have different ratios for $\gamma$ induced recoils.  ER ratios for $\alpha$ interactions and NR ratios are single valued. The numbers listed here represent an error-weighted average of world data as explained in the text. }
\begin{center}
\begin{tabular}{|l|c|}
\hline
 Singlet lifetime & 3.1 $\pm$ 0.7 ns\\
 Triplet lifetime & 24 $\pm$ 1 ns\\
 Singlet/Triplet - ER from  direct excitation ($\gamma$ induced) & $0.17 \pm 0.05$\\
 Singlet/Triplet - ER from recombination ($\gamma$ induced) & $0.8 \pm 0.2$\\
 Singlet/Triplet - ER from both processes ($\alpha$ induced) & $2.3 \pm 0.51$\\
 Singlet/Triplet - NR (neutron induced)& $7.8 \pm 1.5$\\
\hline
\end{tabular}
\end{center}
\label{table:table1}
\end{table} 

The de-excitation lifetime of the dimer  depends on whether it is in the singlet or the triplet state, given by the decays ${}^{1}\Sigma^{+}_{excited} \to {}^{1}\Sigma^{+}_{ground}$ and  ${}^{3}\Sigma^{+}_{excited} \to {}^{1}\Sigma^{+}_{ground}$.  The ratio of the number of excitations in the singlet state to those in the triplet state is dependent on the target medium, recoil type, electric field, and $dE/dx$ \cite{Kubota1979}.   The model includes as its values of the singlet and triplet lifetimes an experimentally determined error-weighted world average, determined by (\ref{eq:avg}),  and empirically determined values for the singlet to triplet ratios to simulate this aspect of the S1 pulse shape.  
\begin{equation}\label{eq:avg}
\bar{\tau} = \sum_{i} \left(\frac{\tau_{i}}{\sigma_{i}}\right) / \sum_{i} \left(\frac{1}{\sigma_{i}} \right);~
\bar{\sigma} = \left [ \sum_{i}\frac{1}{\sigma_{i}} \right ]{}^{-1}.
\end{equation}
Data from \cite{Teymourian:2011rs,morikawa} have the smallest errors and are thus given the most weight, and data from \cite{Kubota1979,Hitachi1982, Akimov:2002kq, Kubota1978, Dawson2005,Hitachi1982,Akimov:2002kq} are also included in the NEST world average weighted by their respective errors.  Table \ref{table:table1} lists all of these parameters for xenon.   The singlet to triplet ratio modeled in NEST is simplified in order to ignore the dependence on electric field and $dE/dx$ because experimental data are lacking in these regards.  

The S1 pulse is also determined by the recombination time, which is inversely proportional to the ionization density. To first order, the Linear Energy Transfer (LET) is proportional to ionization density, so it follows that the recombination time is inversely proportional to the LET \cite{Kubota1979}.   The denser a charge distribution, the easier (and faster) it is for an ionization electron to recombine.  Due to second order effects such as thermal diffusion and atomic line emission and absorption, the LET might not be directly proportional to ionization density, but the NEST framework does not include these effects because such second order effects were not needed for the model to explain observed data \cite{Szydagis:2011tk}.  Simply modeling the recombination time constant as inversely proportional to LET did not fit the experimental data. Hence, we developed a sophisticated model motivated by the recombination probability employed in NEST (Equation 2.6 in \cite{Szydagis:2011tk}):

\begin{equation}
 r = \frac{A \times LET}{1+ B \times LET} + C
 \end{equation}
\noindent where $A$ (0.18) and $B$ (0.42) are parameters of the recombination probability.  Based on this, equation (\ref{eq:recombination}) shows our model for the recombination time constant for the case of zero electric field,

\begin{equation}\label{eq:recombination}
\tau_{r,0} = \hat \tau \times \frac{1+B\times LET}{A \times LET} ns,
\end{equation}

\noindent where $\hat \tau$ is a normalization factor.  The recombination time is modeled as the inverse of the recombination probability.  However, we do not include the offset term ($C$)  because it accounts for geminate recombination, a fast process with a negligible effect on the overall recombination time \cite{Szydagis:2011tk}.  In addition, our model uses an LET that varies stochastically per interaction using Geant4, as opposed to previous models \cite{Ni2006} that have used an average LET.

\subsection{Electron Recoils}
The normalization parameter $\hat \tau$ in (\ref{eq:recombination}) must be extracted from data.  The standard practice is to fit a single exponential to the falling edge of the S1 pulse, a procedure that groups the singlet, triplet, and recombination time into one time parameter.  It is also common to cite 45 ns \cite{Hitachi1982} as the recombination time constant, but this time corresponds only to a 1 MeV electron, ignores the energy dependence of the recombination time, and is actually a single exponential fit encompassing all time constants and not the recombination time.   The approach presented here leads to a more accurate, energy dependent representation of the recombination time constant.

\begin{figure}[h]
\begin{center}
  \includegraphics[width=5.0in]{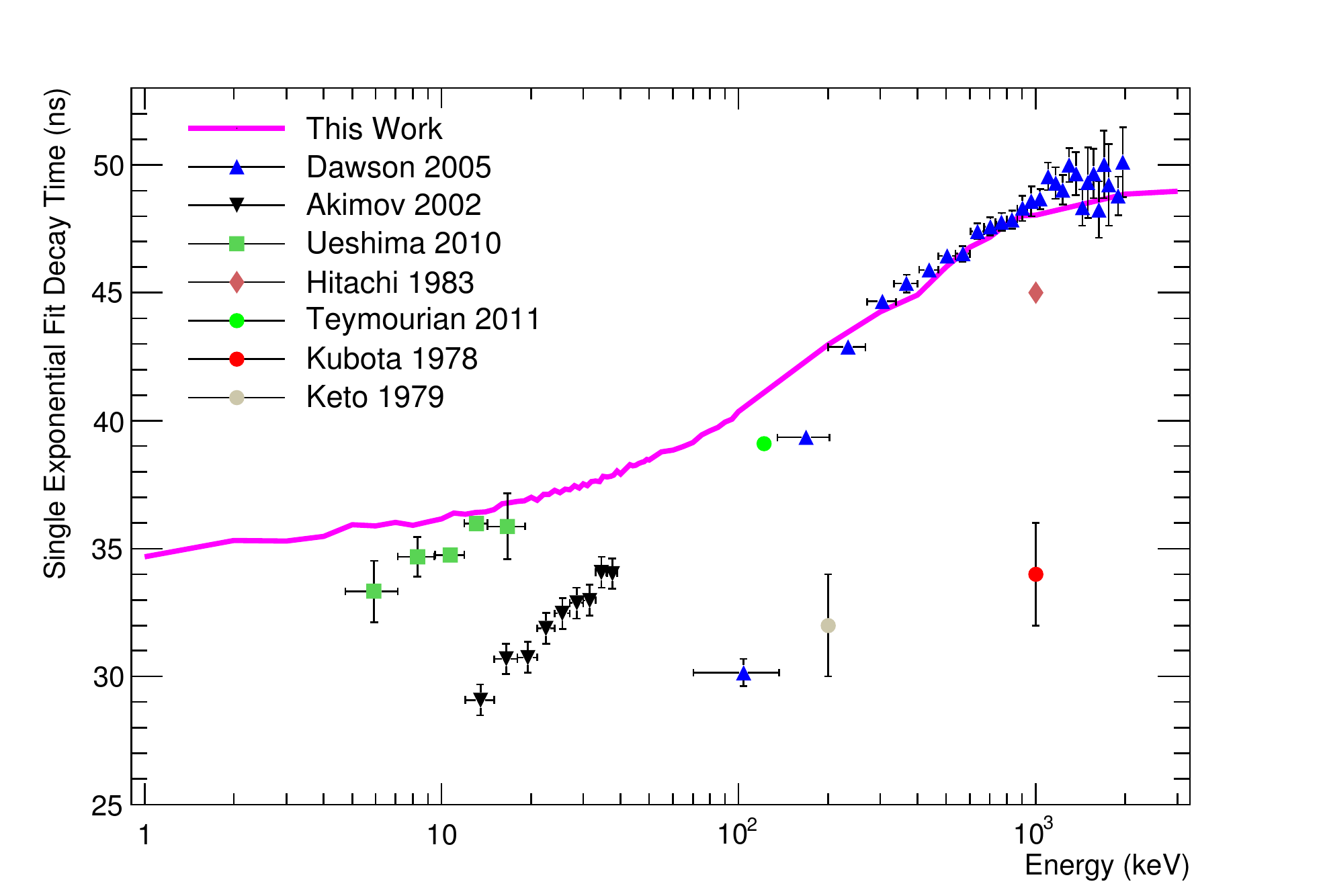}
\end{center}
\caption{A comparison of the single exponential fit to the fall time of the S1 pulse at zero applied electric field for various data sets  and the model presented in this work.  The data were used to minimize the recombination time constant in the model.}
\label{fig:recomProb}
\end{figure}

In our procedure to determine $\hat \tau$, the singlet and triplet lifetimes remain fixed per Table \ref{table:table1}.  We follow an iterative procedure in which initially  $\hat \tau$ is fixed and single-exponential fits are made to the falling edges of simulated S1 pulses as a function of energy.   The extracted single exponential parameter is compared to data from \cite{Dawson2005}, \cite{Akimov:2002kq}, and \cite{XMASS}. Next, $\hat \tau$ is varied and the process is repeated until the error between simulation and data is minimized. For ER events, we determine a value of 3.5 ns for $\hat \tau$.  Figure \ref{fig:recomProb} shows the decay time versus energy as derived from our single exponential fits. Various experimental data are also shown  \cite{Dawson2005, Akimov:2002kq, XMASS, Hitachi1982, Teymourian:2011rs, Kubota1978, 1979} .  As visible in the figure, the data do not agree with one another, so some choices were made.  At high energies, data from Dawson {\it et. al} \cite{Dawson2005} were used in the minimization because it is an extensive and more recent measurement.  At low energies, XMASS \cite{XMASS} was used because  it is a more recent experiment, and hence has better xenon purity. The recombination time model describes the high energy points very well, but is somewhat higher than \cite{XMASS} data points.  This is understandable because those points have larger errors.

\begin{figure}[h]
\begin{center}
 \includegraphics[width=5.0in]{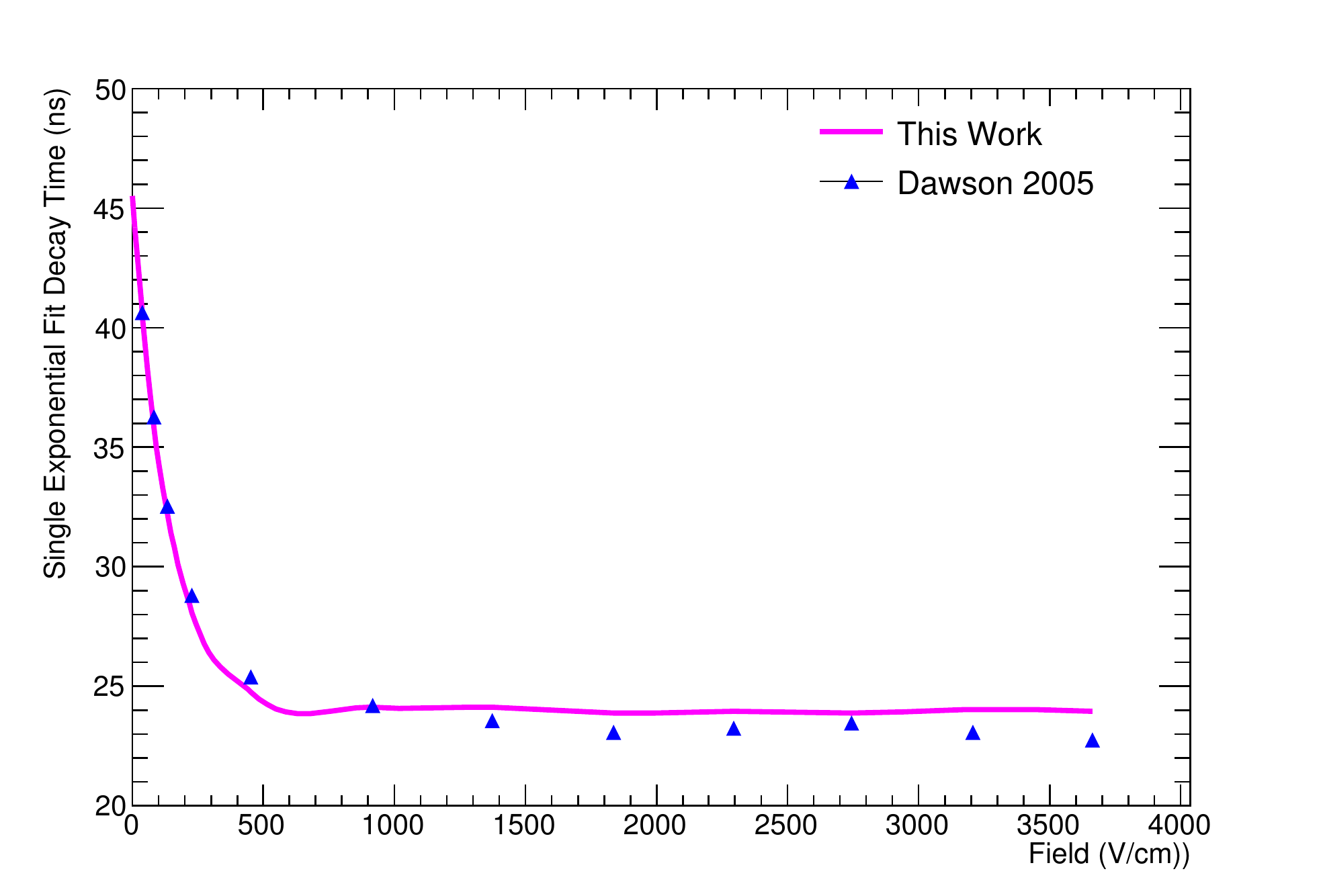}
 \end{center}
\caption{The electric field dependence of the single exponential fit to the fall time of the S1 pulse as the electric field increases for simulated 1.2 MeV gammas compared to similar data from \cite{Dawson2005}.  This decay time is dominated by the recombination time constant which is modeled in this work.}
\label{fig:field}
\end{figure}

The presence of an electric field exponentially quenches recombination.  As the field gets stronger, electrons are more easily stripped away and are less likely to recombine.  Therefore, as the applied electric field goes to infinity the experimentally measured recombination time goes to zero \cite{Dawson2005}, and the triplet decay time constant dominates the time structure of S1 pulse shape.  The electric field dependence is found from an exponential fit to data \cite{Dawson2005}, and (\ref{eq:recombination}) is expanded to,

\begin{equation}
\label{eq:fieldRecomb}
\tau_{r} = \tau_{r,0} \times e^{-0.009E_f},
\end{equation}
\noindent where $E_f$ is the applied electric field in V/cm.  Figure \ref{fig:field} shows the single exponential fit to the fall time of the S1 pulse as electric field increases with (\ref{eq:fieldRecomb}) applied in the model along with the data from \cite{Dawson2005}.  This is the only available data set with a non-zero field, and is at a fixed energy of 1.2 MeV gammas. Until there are more data available, we apply (2.3)  at all energies.

The ratio of the number of singlet states to triplet states is different for direct excitation and recombination processes for electron recoils.  These ratios are measured empirically and included in the model as $0.17 \pm 0.5$ for excitons and $0.8 \pm 0.2$ for recombining electrons \cite{Kubota1979}.

Electron recoil events stemming from alpha interactions are treated differently in the model because they behave differently in the data.  For alpha particles, the recombination time is zero because alphas have a high ionization density. The triplet lifetime will therefore dominate the pulse shape because it is longer than the singlet lifetime.  An approximate fit  to data from \cite{Dawson2005} and \cite{Teymourian:2011rs} yields a singlet to triplet ratio for direct excitation and recombination of $2.3 \pm 0.51$.

\subsection{Nuclear Recoils}
The case of nuclear recoils is different because the high density of ionization results in rapid recombination.  NEST models this negligible recombination time as zero and the pulse shape is determined dominantly by the triplet time constant.  Thus, the only parameter necessary to describe the time evolution of the pulse is the ratio of production of singlet and triplet states. This ratio has been measured to be  $7.8 \pm 1.5$ for direct excitation and recombination \cite{Hitachi1982, Akimov:2002kq}. 

To demonstrate the effectiveness of the S1 pulse shape models, data from a xenon detector from Kwong et al. \cite{Kwong:2009cx} was compared to data from a simulation of the same detector using NEST.  A single exponential was fit to the falling edge of the S1 pulse, and the time constant from the fit to data was compared to that from the simulation for both NR and ER types of events.  These fit decay times are summarized in Table \ref{table:kwong}.

\begin{table}
\caption{The fit time constant resulting from a single exponential fit to the falling edge of S1 pulses from data (Kwong et al.) and a simulation of their detector using the NEST models presented here.  The quoted error bars are the statistical errors from the fit exponential.}
\begin{center}
\begin{tabular}{|c|c|c|}
\hline
 Recoil & Detector Pulse Decay Time (ns)  & Simulated Detector Pulse Decay Time (ns)\\
 \hline
 Electron Recoils & 29.1 $\pm$ 0.3 & 25.3 $\pm$ 0.3\\
 Nuclear Recoils & 29.2 $\pm$ 0.1 & 23.1 $\pm$ 0.1\\ 
 \hline
\end{tabular}
\end{center}
\label{table:kwong}
\end{table} 

\section{S2 Pulse Shape}
The S2 signal is produced in a time projection chamber (TPC) by electrons that escape recombination, are drifted through the detector volume, and are extracted to a higher electric field region, where they produce light via electroluminescence.   The shape of the S2 pulse, for events with long drift times, is roughly Gaussian with characteristic width determined by several parameters: a) the mean free path for electrons to produce photons in xenon gas, b) the drift velocities in the detector volume drift region (DR) and the electroluminescence region (ELR), c) the singlet and triplet state lifetimes, d) the electron diffusion in the DR and the ELR, and e) the trapping of electrons at the interface from DR to ELR.  Each of these effects is incorporated into the model presented here.

\subsection{Mean Free Path}
The number of photons produced by electrons in the ELR is a linear function of electric field per unit density, also known as the reduced field, according to \cite{Monteiro:2007vz}
\begin{equation}\label{eq:meanFreePath}
\frac{n_{ph}}{x} = \left(0.140\frac{E_e}{N}\times 10^{17}-0.474\right)\times N \times 10^{-17}.
\end{equation}
In (\ref{eq:meanFreePath}), $n_{ph}$ is the number of photons produced per ionization electron, $x$ is the distance traveled by the electron in cm, $E_e$ is the electric field given in V/cm and $N$ is the number density of the gas in atoms/cm$^3$.  From (\ref{eq:meanFreePath}), it is clear that the yield $n_{ph}/x$ increases as the reduced field increases.  Alternate treatments formulate this yield in terms of pressure \cite{Monteiro:2007vz}  \cite{Fonseca2004}, however, ignoring temperature dependence implies that they are applicable only to room temperature gas, and not to cold gas or liquid.  Furthermore, light yield changes appreciably in the gaseous phase if it is in the proximity of a liquid phase volume \cite{Fonseca2004}, presumably because of  the presence of micro-droplets of liquid in the gas volume.  The normalization factor of 0.140 has been measured to be as much as 50\% lower in some instances \cite{Monteiro:2007vz}, and a possible explanation for this lies in poor gas purity.  We believe we have a heuristic explanation for this value of  0.140 in (\ref{eq:meanFreePath}), which must carry units of photons/eV.  When inverted, this is $\sim$ 7~eV/photon, which is the energy of the xenon scintillation photons \cite{AprileBook}.  We conclude, therefore, that this particular value suggests there is $\sim 100\%$ efficiency in conversion of electrical energy from the field into scintillation photons.

In the model presented here, to calculate the mean free path of an electron to produce one photon in the ELR, we rearrange (\ref{eq:meanFreePath}) as
\begin{equation}\label{eq:meanFreePath2}
x = \frac{n_{ph}}{N\times 10^{-17}}\left(0.140\frac{E_e}{N}\times 10^{17}-0.474\right)^{-1}.
\end{equation}
This quantity is stochastically applied in Geant4, providing a novel approach for this calculation.  This method is not as detailed as the microphysical approach of \cite{santos}, but it is pragmatic in that it allows for fast and easy simulations that are empirically accurate.

\subsection{Drift Velocity}
The ionization electrons in the DR propagate along the direction of the electric field, and quickly reach terminal velocity ($v_{drift}$) due to collisions with xenon atoms \cite{gushchin1981,Miller:1968zza, Shibamura:1975zz}.  Upon extraction into the ELR, the electric field typically increases.  They collide with xenon atoms, deposit energy, and are re-accelerated.  To model this process, several functions were evaluated for drift velocity, e.g., a sawtooth pattern.  However, these complicated formulations were deemed unnecessary in order for the model to reproduce the data, and a simple, constant drift velocity was implemented.  The Magboltz \cite{Magboltz} package was used to model the drift speed.  We provided a large range of values of detector parameters (electric field and density)  as inputs to Magboltz and tabulated the output drift speeds.  The resulting data were implemented as a look-up table in the simulation package.

In the case of liquid xenon as the DR, there are two established empirical models for drift speed \cite{gushchin1981, Miller:1968zza}, which are both implemented in NEST.  However, the default model is chosen to be \cite{Miller:1968zza} because it is a better match to recent data \cite{Sorensen:2010hv}.  Also, note that NEST implements a linear dependence on temperature, as outlined in \cite{sorensenThesis}. 

\subsection{Exciton Lifetime}
In a manner similar to the S1 process, the S2 pulse in the ELR is generated via the direct excitation and subsequent de-excitation of the medium.  The de-excitation contains a singlet and a triplet component, with characteristic time constants for that medium.  In gaseous xenon, the error-weighted world average singlet decay time constant is $5.88  \pm 5.5$~ns and the triplet decay time constant is $100.1 \pm 7.9$~ns \cite{morikawa, PhysRevA.13.1787, 1979, millet:92, Wenck1980, thornton:133, bonifield:2914, moutard1988, salamero, PhysRevA.9.768}.  The triplet lifetime and singlet lifetime in gas xenon are different than in liquid.  One possible explanation is the vast number of time constants that have been measured with gaseous xenon detectors \cite{AprileBook}.  There is no large consistency between these measurements except for the 100 ns triplet lifetime.

Unlike the case for liquid xenon discussed above, the need to separate the singlet and triplet times for direct excitation and recombination is not important for gaseous xenon. We believe that this is because the drifting electrons are too low in energy to substantially ionize the medium.  Further, as an ansatz for xenon, the model equates the singlet to triplet ratio in gas to that in liquid, an assumption that  successfully reproduces the data.     These two simplified time constants provide a coherent model of this aspect of the scintillation physics. 

Figure \ref{dusty}, bottom, shows a simulation of a typical S2 pulse as various physical processes are included in the model.  If only the ionization electron drift is modeled, the shape of the pulse is a square wave.  Adding in the singlet and triplet lifetimes discussed in this section changes the pulse shape to one that resembles a shark fin \cite{Koehler}.  The red curve shows the result of including the drift speed in the DR and ELR, the mean free path of an electron to produce a photon, and the singlet and triplet decay time constants in the model.  This does not yet provide a complete picture of the S2 pulse shape, as will be discussed below.

\subsection{Diffusion}
Ionization electrons that drift in a TPC will experience diffusion in three dimensions, modeled here as transverse (perpendicular to the direction of drift) and longitudinal (parallel to the direction of drift) diffusion.  The resulting electron cloud is  an ellipsoid with minor axis along the direction of drift.  The diffusion can be described as \cite{Sorensen:2011qs}
\begin{equation}\label{eq:diffusion}
\sigma_L = \sqrt{2D_L \Delta t};~
\sigma_T = \sqrt{2D_T \Delta t};~
~\Delta t = \Delta z/v_{drift}
\end{equation}
where, $\sigma_L$ and $D_L$ ($\sigma_T$ and $D_T$) are the longitudinal (transverse) diffusion width and constant respectively, $\Delta t$ is the drift time, and $\Delta z$ is the longitudinal drift distance .  The diffusion constants are modeled as power law fits \cite{Doke1976} dependent on the drift electric field, thus eliminating diffusion as a free parameter, unlike the procedure followed in \cite{Sorensen:2011qs}.  

The DR in a two-phase xenon TPC is liquid, and both transverse and longitudinal diffusion are considered in the model.  Transversal diffusion has no first-order effect on the shape of the S2 pulse, which is mostly determined by the temporal (hence longitudinal) distribution of electrons.  In fact, the transverse diffusion is smaller than the position resolution of a typical TPC \cite{Aprile2010, Aprile2010a, Akerib:2012ak}.  An increase in the density of photodetectors will, of course, improve the transverse resolution and this effect could become more prominent.  The longitudinal diffusion in liquid xenon is an order of magnitude smaller than transverse, but it has a critical effect on the S2 shape because it dictates the photon time of arrival at the photodetectors.  

In our implementation in Geant4, diffusion and drift are modeled separately as follows.  A cloud of ionization electrons is created in the DR.  Depending on $v_{drift}$ and $\Delta z$, the diffusion widening of the cloud is computed and applied based on (\ref{eq:diffusion}).  This ellipsoidal electron cloud is then drifted  to the ELR, a process which includes loss of electrons due to impurities.  The effective impurity concentration is provided by the user in terms of a mean absorption length.  

Both longitudinal and transverse diffusion in gas in the ELR have a negligible effect on the S2 pulse shape as long as the $\Delta z$ of the event is large.  The following reasons describe why: a) the atomic density is much lower, b) the electric field in the ELR is much higher leading to a smaller value of $\Delta t$ in (\ref{eq:diffusion}), c) the drift distance in the ELR is smaller than in the DR and is identical for each event.  In Figure \ref{dusty}, bottom, the magenta curve shows the resulting S2 simulated pulse shape when the gas diffusion constants are added to the physical processes included in the model.  It is evident that this has a negligible effect on the shape.  The effects on the S2 shape when the transverse and longitudinal diffusion in the liquid are included are shown in black and green, respectively.  The effect from the longitudinal diffusion in liquid has the largest influence on the pulse shape.

\subsection{Electron Trapping Time}
The electron extraction time for two-phase TPCs is modeled as an exponential distribution, where the characteristic time constant is derived from a quantum tunneling probability. Electrons drift to the interface between the ELR and DR where they must overcome the work function barrier, assisted by the extraction field.  The emitted electron current density  $J$ $(Am^{-2})$ at the interface is given by the Richardson-Dushman equation and is related to the temperature, $T$ and the work function $W$ as $J = AT^2e^{-W/kT}$

\begin{figure}[h]
\begin{center}
\includegraphics[width=4.5in]{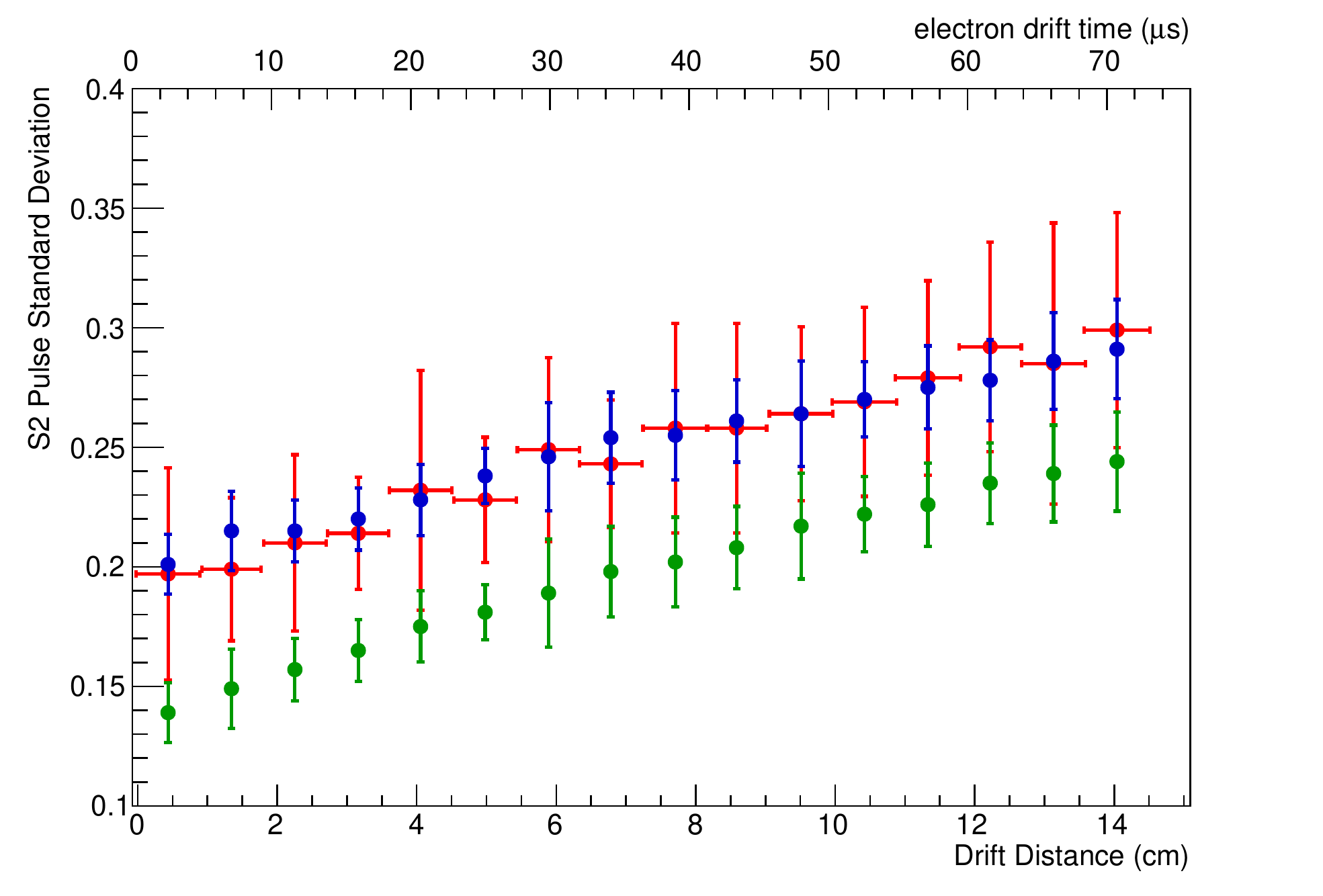}
\includegraphics[width=4.5in]{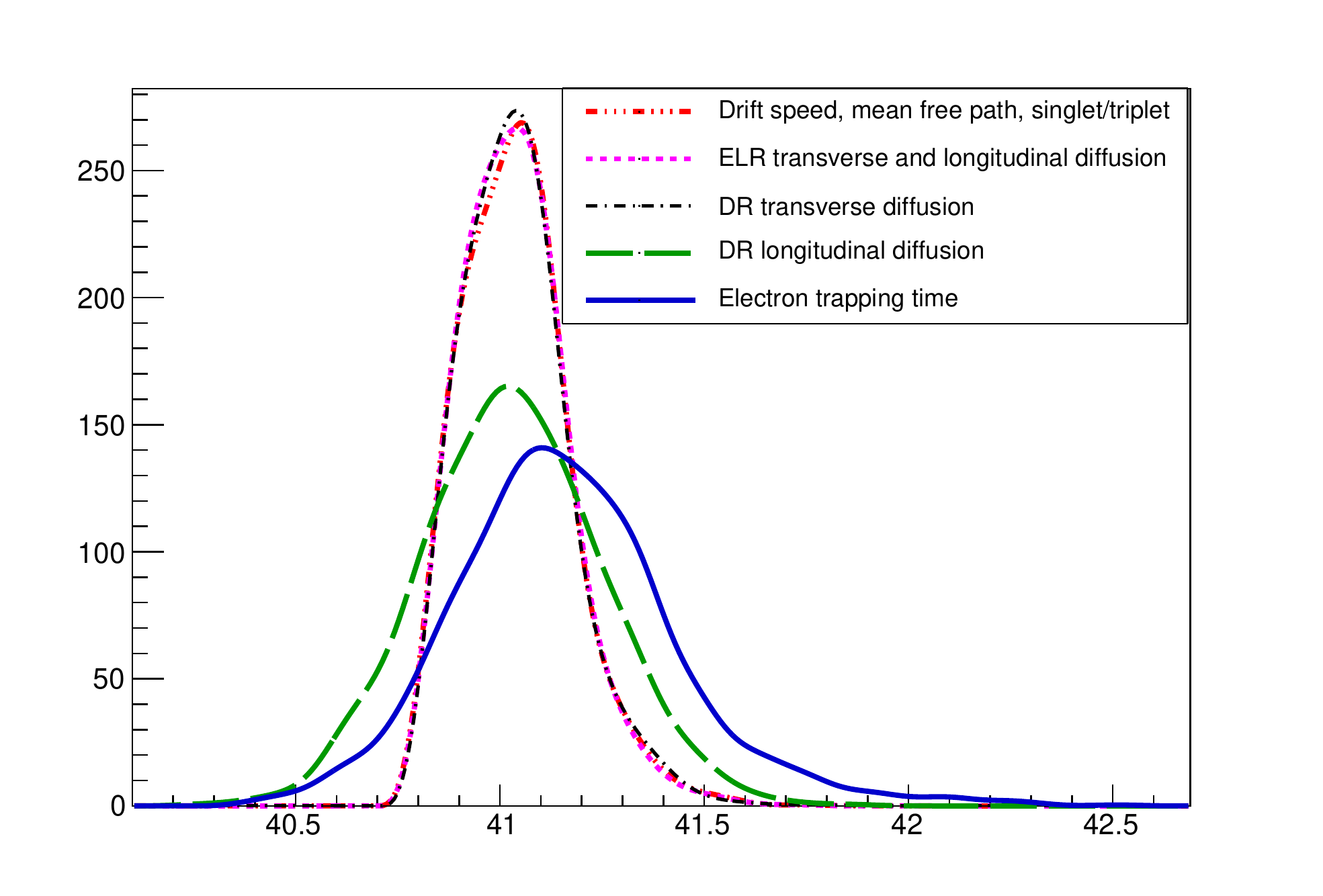}
\end{center}
\caption{{\it Top:} The fitted Gaussian width of an S2 pulse as a function of drift for data from Xenon10 (red) \cite{Sorensen:2011qs} compared to a toy model of that detector incorporating the model presented in this work (blue).  The error bars in the simulated work represent the width on the distribution of the Gaussian fit widths of the S2 pulses, due to the stochastic variation on the number of ionization electrons as provided by NEST.  In the data, the error bars are a combination of this width as well as the uncertainty in the fit.  The green points show the fitted width due to effects derived from first principles.  Inclusion of the electron trapping time as a free parameter allows for excellent agreement between model and data.  {\it Bottom:} The simulation of an average S2 pulse demonstrating the various components of the model.  The red curve includes the drift speed in both the DR and ELR, the mean free path of an electron to produce a photon, and the singlet and triplet ratios and time constants.  The magenta curve adds both longitudinal and transverse diffusion in the ER.  The black curve adds the transverse diffusion affect in the DR.  The green curve adds longitudinal diffusion in the DR.  Finally the blue curve adds in the electron trapping time and shows the simulated S2 pulse shape.}
\label{dusty}
\end{figure}

To study this effect and find the time parameter that best reproduces available data, a simplified model of the XENON10 detector \cite{Angle2008} was built in Geant4.  The parameters used in this toy simulation were: a) in liquid xenon, applied field was $0.73 kV/cm$ and the drift velocity was 1.89 $mm/\mu s$, b) in gas, the electric field was $12 kV/cm$ and the drift speed was 8 $mm/\mu s$ \cite{sorensenThesis}.  S2 events were simulated as a function of drift for nuclear recoil energies in the 5-50 keV range, with the number of ionization electrons provided by NEST  \cite{Szydagis:2011tk}.  The number of photons per ionization electron in the ELR was determined by (\ref{eq:meanFreePath2}).  A Gaussian was fit to each generated S2 pulse, and the width was compared to data \cite{Sorensen:2011qs}.  The results of this study are shown in Figure \ref{dusty}, top.  The red points are the Xenon10 data, and the green points are derived from the toy model, with physically motivated values for the parameters mentioned in the previous sections and no electron trapping time effect added.  As can be clearly seen, this simulation produces S2 pulse widths that are systematically lower than the measurements.

Subsequently, the electron trapping time was tuned as the only free parameter until the model best fit this data.  This value was found to be 140 ns, which is comparable to the expected value of $\sim$100 ns\cite{AprileBook}.  Electrons are drifted through the DR until they reach the phase transition surface.  The electrons are emitted from the surface following an exponential distribution with the electron trapping time as the time constant. Figure \ref{dusty}, top, also shows the blue points, which are the simulated pulse widths after the electron trapping time is included.  It is evident that this parameter is all that is necessary to accurately simulate the correct pulses.  In fact,  we find no need to add an extra constant width in quadrature, as was done previously \cite{Sorensen:2011qs}.  Using a Gaussian to describe the pulse shape of the S2 assumes that the S2 is inherently symmetric, however, it has been shown that this functional form can accurately fit data from Xenon10 and Xenon100 \cite{Sorensen:2011qs}.  Other effects on the pulse shape may exist however we assume these effects are subdominant because the electron trapping time, with a value approximately as expected, allows the model to reproduce the data.

In Figure \ref{dusty}, top, the error bars in the simulated work represent the width on the distribution of the Gaussian fit widths of the S2 pulses, due to the stochastic variation on the number of ionization electrons as provided by NEST.  In the data, the error bars are caused by a combination of this width, the uncertainty in the fit, and the widths introduced by other detector parameters such as data acquisition, PMT response, and analysis efforts.  Because the data contains additional sources that act to broaden the pulse shapes, it is reasonable to assume the error bars will be larger than in the model which does not include these additional detector-specific effects.  It should be noted that these detector-specific parameters should not affect the mean of the distribution, and this is clearly evident in the figure.  The S2 width as it compares to drift distance is important in the search for low mass dark matter particles \cite{Sorensen2010b}.  The blue curve in Figure \ref{dusty}, bottom, shows the model of the final S2 pulse shape including the electron trapping time.

\section{Conclusion}
In this paper, we have presented a comprehensive model for describing the characteristic pulse shapes of signals generated  in xenon detectors.  For the S1 pulse, effects such as the ratio of singlet and triplet dimer state populations, their corresponding decay times, and the recombination time were incorporated in the simulation.  For S2 pulses, a multitude of parameters, namely, the ionization electron mean free path in gas, the drift velocity, singlet and triplet decay times, diffusion constants, and the electron trapping time, have been implemented. The results have been shown to agree with various measurements.  We believe that this is a complete and unique package that  realistically describes expected pulse shapes for two-phase xenon TPCs.

\newpage

\acknowledgments
This work was supported by U.S. Department of Energy grant DE-FG02-91ER40674 at the University of California, Davis, as well as supported by DOE grant DE-NA0000979, which funds the seven universities involved in the Nuclear Science and Security Consortium. 

%\newpage

%\begin{thebibliography}{9}
\bibliography{XenonPaper}
\bibliographystyle{plain}
  
%\end{thebibliography}

\end{document}